\documentclass[twocolumn]{aastex7}

\usepackage{graphicx}
\usepackage{amsmath}
\usepackage{hyperref}
\usepackage{longtable}
\usepackage{booktabs}

\newcommand{\te}{t_{\rm E}}
\newcommand{\thetae}{\theta_{\rm E}}

\newcommand{\pie}{\pi_{\rm E}}

\newcommand{\dl}{D_{\rm L}}

\def\e{{\rm E}}

\begin{document}

\title{Four Giant Planets from 2024 KMTNet Microlensing Campaign }
\shorttitle{Four Giant Planets from 2024 KMTNet}


\author{Cheongho Han}
\affiliation{Department of Physics, Chungbuk National University, Cheongju 28644, Republic of Korea}
\email{cheongho@astroph.chungbuk.ac.kr}
\author{Andrzej Udalski} 
\affiliation{Astronomical Observatory, University of Warsaw, Al.~Ujazdowskie 4, 00-478 Warszawa, Poland}
\email{udalski@astrouw.edu.pl} 
\author{Ian A. Bond}
\affiliation{School of Mathematical and Computational Sciences, Massey University, Auckland 0745, New Zealand}
\email{i.a.bond@massey.ac.nz}
\author{Chung-Uk Lee$^*$}
\affiliation{Korea Astronomy and Space Science Institute, Daejon 34055, Republic of Korea}
\email{leecu@kasi.re.kr}
\author{Jiyuan Zhang}
\affiliation{Department of Astronomy, Tsinghua University, Beijing 100084, China}
\email{zhangjy22@mails.tsinghua.edu.cn}
\collaboration{5}{(Leading authors)}
\author{Michael D. Albrow}   
\affiliation{University of Canterbury, Department of Physics and Astronomy, Private Bag 4800, Christchurch 8020, New Zealand}
\email{michael.albrow@canterbury.ac.nz}
\author{Sun-Ju Chung}
\affiliation{Korea Astronomy and Space Science Institute, Daejon 34055, Republic of Korea}
\email{sjchung@kasi.re.kr}
\author{Andrew Gould}
\affiliation{Department of Astronomy, Ohio State University, 140 West 18th Ave., Columbus, OH 43210, USA}
\email{gould.34@osu.edu}
\author{Youn Kil Jung}
\affiliation{Korea Astronomy and Space Science Institute, Daejon 34055, Republic of Korea}
\affiliation{University of Science and Technology, Daejeon 34113, Republic of Korea}
\email{younkil21@gmail.com}
\author{Kyu-Ha~Hwang}
\affiliation{Korea Astronomy and Space Science Institute, Daejon 34055, Republic of Korea}
\email{kyuha@kasi.re.kr}
\author{Yoon-Hyun Ryu}
\affiliation{Korea Astronomy and Space Science Institute, Daejon 34055, Republic of Korea}
\email{yhryu@kasi.re.kr}
\author{Yossi Shvartzvald}
\affiliation{Department of Particle Physics and Astrophysics, Weizmann Institute of Science, Rehovot 76100, Israel}
\email{yossishv@gmail.com}
\author{In-Gu Shin}
\affiliation{Department of Astronomy, Westlake University, Hangzhou 310030, Zhejiang Province, China}
\email{ingushin@gmail.com}
\author{Jennifer C. Yee}
\affiliation{Center for Astrophysics $|$ Harvard \& Smithsonian 60 Garden St., Cambridge, MA 02138, USA}
\email{jyee@cfa.harvard.edu}
\author{Weicheng Zang}
\affiliation{Department of Astronomy, Westlake University, Hangzhou 310030, Zhejiang Province, China}
\email{zangweicheng@westlake.edu.cn}
\author{Hongjing Yang}
\affiliation{Department of Astronomy, Tsinghua University, Beijing 100084, China}
\affiliation{Department of Astronomy, Westlake University, Hangzhou 310030, Zhejiang Province, China}
\email{yang-hj19@mails.tsinghua.edu.cn}
\author{Doeon Kim}
\affiliation{Department of Physics, Chungbuk National University, Cheongju 28644, Republic of Korea}
\email{qso21@hanmail.net}
\author{Dong-Jin Kim}
\affiliation{Korea Astronomy and Space Science Institute, Daejon 34055, Republic of Korea}
\email{keaton03@kasi.re.kr}
\author{Byeong-Gon Park}
\affiliation{Korea Astronomy and Space Science Institute, Daejon 34055, Republic of Korea}
\email{bgpark@kasi.re.kr}
\collaboration{14}{(KMTNet Collaboration)}
\author{Przemek Mr{\'o}z}
\affiliation{Astronomical Observatory, University of Warsaw, Al.~Ujazdowskie 4, 00-478 Warszawa, Poland}
\email{pmroz@astrouw.edu.pl}
\author{Micha{\l} K. Szyma{\'n}ski}
\affiliation{Astronomical Observatory, University of Warsaw, Al.~Ujazdowskie 4, 00-478 Warszawa, Poland}
\email{msz@astrouw.edu.pl}
\author{Jan Skowron}
\affiliation{Astronomical Observatory, University of Warsaw, Al.~Ujazdowskie 4, 00-478 Warszawa, Poland}
\email{jskowron@astrouw.edu.pl}
\author{Rados{\l}aw Poleski} 
\affiliation{Astronomical Observatory, University of Warsaw, Al.~Ujazdowskie 4, 00-478 Warszawa, Poland}
\email{radek.poleski@gmail.co}
\author{Igor Soszy{\'n}ski}
\affiliation{Astronomical Observatory, University of Warsaw, Al.~Ujazdowskie 4, 00-478 Warszawa, Poland}
\email{soszynsk@astrouw.edu.pl}
\author{Pawe{\l} Pietrukowicz}
\affiliation{Astronomical Observatory, University of Warsaw, Al.~Ujazdowskie 4, 00-478 Warszawa, Poland}
\email{pietruk@astrouw.edu.pl}
\author{Szymon Koz{\l}owski} 
\affiliation{Astronomical Observatory, University of Warsaw, Al.~Ujazdowskie 4, 00-478 Warszawa, Poland}
\email{simkoz@astrouw.edu.pl}
\author{Krzysztof A. Rybicki}
\affiliation{Astronomical Observatory, University of Warsaw, Al.~Ujazdowskie 4, 00-478 Warszawa, Poland}
\affiliation{Department of Particle Physics and Astrophysics, Weizmann Institute of Science, Rehovot 76100, Israel}
\email{krybicki@astrouw.edu.pl}
\author{Patryk Iwanek}
\affiliation{Astronomical Observatory, University of Warsaw, Al.~Ujazdowskie 4, 00-478 Warszawa, Poland}
\email{piwanek@astrouw.edu.pl}
\author{Krzysztof Ulaczyk}
\affiliation{Department of Physics, University of Warwick, Gibbet Hill Road, Coventry, CV4 7AL, UK}
\email{kulaczyk@astrouw.edu.pl}
\author{Marcin Wrona}
\affiliation{Astronomical Observatory, University of Warsaw, Al.~Ujazdowskie 4, 00-478 Warszawa, Poland}
\affiliation{Villanova University, Department of Astrophysics and Planetary Sciences, 800 Lancaster Ave., Villanova, PA 19085, USA}
\email{mwrona@astrouw.edu.pl}
\author{Mariusz Gromadzki}          
\affiliation{Astronomical Observatory, University of Warsaw, Al.~Ujazdowskie 4, 00-478 Warszawa, Poland}
\email{marg@astrouw.edu.pl}
\author{Mateusz J. Mr{\'o}z} 
\affiliation{Astronomical Observatory, University of Warsaw, Al.~Ujazdowskie 4, 00-478 Warszawa, Poland}
\email{mmroz@astrouw.edu.pl}
\collaboration{100}{(The OGLE Team)}
\author{Fumio Abe}
\affiliation{Institute for Space-Earth Environmental Research, Nagoya University, Nagoya 464-8601, Japan}
\email{abe@isee.nagoya-u.ac.jp}
\author{David P. Bennett}
\affiliation{Code 667, NASA Goddard Space Flight Center, Greenbelt, MD 20771, USA}
\affiliation{Department of Astronomy, University of Maryland, College Park, MD 20742, USA}
\email{bennett.moa@gmail.com}
\author{Aparna Bhattacharya}
\affiliation{Code 667, NASA Goddard Space Flight Center, Greenbelt, MD 20771, USA}
\affiliation{Department of Astronomy, University of Maryland, College Park, MD 20742, USA}
\email{aparna.bhattacharya@nasa.gov}
\author{Ryusei Hamada}
\affiliation{Department of Earth and Space Science, Graduate School of Science, Osaka University, Toyonaka, Osaka 560-0043, Japan}
\email{hryusei@iral.ess.sci.osaka-u.ac.jp}
\author{Stela Ishitani Silva}  
\affiliation{Code 667, NASA Goddard Space Flight Center, Greenbelt, MD 20771, USA}
\affiliation{Department of Physics, The Catholic University of America, Washington, DC 20064, USA}
\email{ishitanisilva@cua.edu}
\author{Yuki Hirao}
\affiliation{Institute of Astronomy, Graduate School of Science, The University of Tokyo, 2-21-1 Osawa, Mitaka, Tokyo 181-0015, Japan}
\email{hirao@ioa.s.u-tokyo.ac.jp}
\author{Asahi Idei}
\affiliation{Department of Earth and Space Science, Graduate School of Science, Osaka University, Toyonaka, Osaka 560-0043, Japan}
\email{idei@iral.ess.sci.osaka-u.ac.jp}
\author{Shota Miyazaki}
\affiliation{Institute of Space and Astronautical Science, Japan Aerospace Exploration Agency, 3-1-1 Yoshinodai, Chuo, Sagamihara, Kanagawa 252-5210, Japan}
\email{miyazaki@ir.isas.jaxa.jp}
\author{Yasushi Muraki}
\affiliation{Institute for Space-Earth Environmental Research, Nagoya University, Nagoya 464-8601, Japan}
\email{muraki@isee.nagoya-u.ac.jp}
\author{Tutumi Nagai}
\affiliation{Department of Earth and Space Science, Graduate School of Science, Osaka University, Toyonaka, Osaka 560-0043, Japan}
\email{nagai@iral.ess.sci.osaka-u.ac.jp}
\author{Kansuke Nunota}
\affiliation{Department of Earth and Space Science, Graduate School of Science, Osaka University, Toyonaka, Osaka 560-0043, Japan}
\email{unota@iral.ess.sci.osaka-u.ac.jp}
\author{Greg Olmschenk}
\affiliation{Code 667, NASA Goddard Space Flight Center, Greenbelt, MD 20771, USA}
\email{greg@olmschenk.com}
\author{Cl{\'e}ment Ranc}
\affiliation{Sorbonne Universit\'e, CNRS, UMR 7095, Institut d'Astrophysique de Paris, 98 bis bd Arago, 75014 Paris, France}
\email{ranc@iap.fr}
\author{Nicholas J. Rattenbury}
\affiliation{Department of Physics, University of Auckland, Private Bag 92019, Auckland, New Zealand}
\email{n.rattenbury@auckland.ac.nz}
\author{Yuki Satoh}
\affiliation{College of Science and Engineering, Kanto Gakuin University, Yokohama, Kanagawa 236-8501, Japan}
\email{yukisato@kanto-gakuin.ac.jp}
\author{Takahiro Sumi}
\affiliation{Department of Earth and Space Science, Graduate School of Science, Osaka University, Toyonaka, Osaka 560-0043, Japan}
\email{sumi@ess.sci.osaka-u.ac.jp}
\author{Daisuke Suzuki}
\affiliation{Department of Earth and Space Science, Graduate School of Science, Osaka University, Toyonaka, Osaka 560-0043, Japan}
\email{dsuzuki@ir.isas.jaxa.jp}
\author{Takuto Tamaoki}
\affiliation{Department of Earth and Space Science, Graduate School of Science, Osaka University, Toyonaka, Osaka 560-0043, Japan}
\email{tamaoki@iral.ess.sci.osaka-u.ac.jp}
\author{Sean K. Terry}
\affiliation{Code 667, NASA Goddard Space Flight Center, Greenbelt, MD 20771, USA}
\affiliation{Department of Astronomy, University of Maryland, College Park, MD 20742, USA}
\email{skterry@umd.edu}
\author{Paul J. Tristram}
\affiliation{University of Canterbury Mt.~John Observatory, P.O. Box 56, Lake Tekapo 8770, New Zealand}
\email{tristram.p@gmail.com}
\author{Aikaterini Vandorou}
\affiliation{Code 667, NASA Goddard Space Flight Center, Greenbelt, MD 20771, USA}
\affiliation{Department of Astronomy, University of Maryland, College Park, MD 20742, USA}
\email{aikaterini.vandorou@utas.edu.au}
\author{Hibiki Yama}
\affiliation{Department of Earth and Space Science, Graduate School of Science, Osaka University, Toyonaka, Osaka 560-0043, Japan}
\email{yama@iral.ess.sci.osaka-u.ac.jp}
\collaboration{100}{(The MOA Collaboration)}
\author{Yuchen Tang}
\affiliation{Department of Astronomy, Westlake University, Hangzhou 310030, Zhejiang Province, China}
\email{yuchen.tanggg@qq.com}
\author{Yunyi Tang}
\affiliation{Department of Astronomy, Tsinghua University, Beijing 100084, China}
\email{tangyy19@mails.tsinghua.edu.cn}
\author{Shude Mao}
\affiliation{Department of Astronomy, Tsinghua University, Beijing 100084, China}
\email{shude.mao@gmail.com}
\author{Dan Maoz}
\affiliation{School of Physics and Astronomy, Tel-Aviv University, Tel-Aviv 6997801, Israel}
\email{maoz@astro.tau.ac.il}
\author{Wei Zhu}
\affiliation{Department of Astronomy, Tsinghua University, Beijing 100084, China}
\email{weizhu@mail.tsinghua.edu.cn}
\collaboration{100}{(The LCO Team)}
\correspondingauthor{\texttt{cheongho@astroph.chungbuk.ac.kr}}
\correspondingauthor{\texttt{leecu@kasi.re.kr}}

\begin{abstract}
In this work, we present analyses of four newly discovered planetary microlensing 
events from the 2024 KMTNet survey season: KMT-2024-BLG-0176, KMT-2024-BLG-0349, 
KMT-2024-BLG-1870, and KMT-2024-BLG-2087.  In each case, the planetary nature was 
revealed through distinct types of anomalies in the lensing light curves: a positive 
bump near the peak for KMT-2024-BLG-0176, an asymmetric peak for KMT-2024-BLG-0349, 
a short-duration central dip for KMT-2024-BLG-1870, and a caustic-crossing feature 
for KMT-2024-BLG-2087. Detailed modeling of the light curves confirms that these 
anomalies are produced by planetary companions with planet-to-host mass ratios in 
the range of $(1.5\text{--}17.9)\times 10^{-3}$.  Despite the diversity of signal 
morphologies, all planets detected in these events are giant planets with masses 
comparable to or exceeding that of Jupiter in the Solar System. Each planet orbits 
a host star less massive than the Sun, emphasizing the strength of microlensing in 
uncovering planetary systems around low-mass stellar hosts.  
\end{abstract}


\section{Introduction} \label{sec:one}

Gravitational microlensing is a powerful technique for detecting exoplanets that are 
difficult or impossible to discover using other major methods, such as radial velocity 
and transits. These traditional techniques are most sensitive to planets in close-in 
orbits around relatively bright host stars, resulting in observational biases toward 
certain types of planetary systems. In contrast, microlensing is uniquely sensitive 
to planets at wider separations from their hosts, including those beyond the snow 
line, where planet formation processes are believed to differ substantially. Moreover, 
because microlensing does not rely on detecting light from the host star, it can reveal 
planets orbiting intrinsically faint low-mass stars, which are the most common stellar 
population in the Galaxy.

\begin{table*}[t]
\centering
\caption{Event coordinates and ID correspondence.  \label{table:one}}
\begin{tabular}{llllllllllcc}
\hline\hline
\multicolumn{1}{c}{Event}                     &
\multicolumn{1}{c}{(RA, DEC)$_{\rm J2000}$}   &
\multicolumn{1}{c}{$(l, b)$}                  &
\multicolumn{1}{c}{Other data}                \\
\hline
KMT-2024-BLG-0176  & (17:46:07.79, -34:12:38.09) &  (-4$^\circ$\hskip-2pt.4506, -2$^\circ$\hskip-2pt.8354)    &  LCOS                                   \\
KMT-2024-BLG-0349  & (17:25:29.17, -30:14:49.27) &  (-3$^\circ$\hskip-2pt.4670,  2$^\circ$\hskip-2pt.9769)    &  OGLE-2024-BLG-0595, LCOA               \\
KMT-2024-BLG-1870  & (18:09:51.68, -26:52:58.40) &  ( 4$^\circ$\hskip-2pt.4587, -3$^\circ$\hskip-2pt.6099)    &  OGLE-2024-BLG-1171, MOA-2024-BLG-206   \\
KMT-2024-BLG-2087  & (17:53:48.01, -29:15:47.48) &  ( 0$^\circ$\hskip-2pt.6374, -1$^\circ$\hskip-2pt.7011)    &  MOA                                    \\
\hline
\end{tabular}
\end{table*}

By probing a population of planets largely inaccessible to other detection 
methods, microlensing plays a vital role in developing a more complete and 
less biased understanding of planetary demographics across the Galaxy.  
Building on this strength, \citet{Zang2025} recently analyzed microlensing 
planets discovered by high-cadence surveys between the 2016 and 2019 seasons 
to characterize the distribution of planet-to-host mass ratios for wide-orbit 
planets.  Their study found that, on average, there are about 0.35 super-Earths 
per star located at orbital distances comparable to that of Jupiter. Furthermore, 
the mass-ratio distribution was found to be bimodal, with peaks corresponding 
to super-Earths and to gas giants like Jupiter and Saturn. This pattern offers 
important empirical constraints on theories of planet formation and migration, 
particularly in the cold outer regions of planetary systems.

A statistically meaningful demographic analysis requires a sufficiently large and 
representative sample of detected planets. In microlensing, such a sample has been 
made possible primarily by three high-cadence survey programs: the Optical Gravitational 
Lensing Experiment \citep[OGLE;][]{Udalski2015}, the Microlensing Observations in 
Astrophysics project \citep[MOA;][]{Bond2001, Sumi2003}, and the Korea Microlensing 
Telescope Network \citep[KMTNet;][]{Kim2016}. These surveys continuously monitor 
dense stellar fields toward the Galactic bulge with high cadence and wide sky coverage, 
detecting up to 4,000 microlensing events annually.  Among these events, about 10\% 
show deviations from the smooth, symmetric light curves characteristic of single-lens 
single-source (1L1S) configurations. Such anomalies can arise from a variety of causes, 
including binary lenses and binary sources. Notably, about 10\% of the anomalous events 
are attributable to planetary companions to the lens, implying that microlensing surveys 
detect roughly 30 planets per year \citep{Gould2022}.

Some microlensing planets are published individually because of their distinctive 
characteristics or the unique insights they offer into planetary formation and 
evolution. Notable examples include the Earth-mass planets OGLE-2016-BLG-1195Lb 
\citep{Bond2017, Shvartzvald2017} and OGLE-2013-BLG-0341Lb \citep{Gould2014}, as 
well as the free-floating planet candidates OGLE-2016-BLG-1540L \citep{Mroz2018} 
and KMT-2017-BLG-2820 \citep{Ryu2021}. Other planets with similar microlensing 
signal characteristics are often presented collectively. These include planets 
detected through major- or minor-image perturbations \citep{Han2024b, Han2024a}, 
low-mass-ratio planets producing dip-like anomalies \citep{Han2025a}, planets 
identified via non-caustic-crossing \citep{Han2021a, Han2023} or resonant-caustic 
channels \citep{Han2021b}, and events exhibiting weak caustic-crossing features 
\citep{Han2025c}.

To enable robust demographic studies, microlensing detections must be published 
systematically so that a complete, unbiased dataset can be assembled. Without 
comprehensive reporting of routine events, statistical inference is susceptible 
to publication bias, skewed toward unusual or striking planets, thereby undermining 
population-level conclusions based on mass-ratio and separation distributions. 
Yet, despite the importance of sample completeness, some planets remain unpublished, 
limiting their contribution to statistical analyses. Ensuring that all detected 
planets are documented in the literature is therefore essential for reliable 
demographic studies.

In this context, the KMTNet team has pursued a systematic analysis of survey 
data since 2016. These efforts have steadily expanded the published sample: 
\citet{Hwang2022} analyzed prime-field data from the 2018–2019 seasons and 
reported six planets; \citet{Ryu2024} and \citet{Gould2022} examined the 2017 
and 2018 prime fields, reporting three and eight planets, respectively; 
\citet{Jung2022, Jung2023} presented six and five planets from the 2018 and 
2019 subprime fields; \citet{Zang2022} reported three planets from the 2019 
season; and \citet{Zang2023} identified seven very low mass-ratio planets from 
four years of KMTNet data (2016–2019). Additional contributions include five 
planets and one candidate from the 2016 prime fields \citep{Shin2023}, four 
more from the 2016 subprime fields \citep{Shin2024}, and four confirmed planets 
plus one candidate from the 2017 subprime fields \citep{Gui2024}. Most recently, 
\citet{Han2025b} added four planets identified in events with faint source stars, 
with the explicit aim of incorporating them into future statistical analyses.

The present work continues this systematic approach by presenting four new 
microlensing planets detected during the 2024 KMTNet season. These planets 
are all giants orbiting sub-solar-mass hosts, located in both the Galactic 
disk and bulge.  The planets reveal their presence through signals with diverse 
morphologies, such as a positive bump, an asymmetric peak, a dip, and a caustic 
crossing, thereby demonstrating the variety of microlensing properties.  By 
publishing these events together as part of the 2024 KMTNet release, we ensure 
that they are properly documented, accessible, and available for both demographic 
studies and future applications such as algorithmic training and survey design.

\section{Discoveries and data} \label{sec:two}

The planets presented in this study were initially identified from anomalous 
features in the light curves of microlensing events detected by the KMTNet 
survey. These anomalies were detected through a combination of the semi-automatic 
AnomalyFinder algorithm \citep{Zang2021} and visual inspection. Preliminary 
analyses were performed using online data processed by the automated photometry 
pipeline. For events with signals suggestive of planetary origin, we conducted 
detailed modeling based on refined photometry derived from optimized re-reduction 
of the data. We also incorporated additional observations from other microlensing 
surveys, including OGLE and MOA. This process led to the identification of four
planetary events: KMT-2024-BLG-0176,KMT-2024-BLG-0349, KMT-2024-BLG-1870, and 
KMT-2024-BLG-2087.

Table~\ref{table:one} presents the equatorial and Galactic coordinates of the events. 
Among these events, KMT-2024-BLG-0349 was also observed by OGLE, KMT-2024-BLG-2087 
was additionally monitored by MOA, and KMT-2024-BLG-1870 was observed by both OGLE 
and MOA.  For events monitored by multiple surveys, the table includes the event 
identifiers assigned by each respective group.  In addition to the survey data, 
follow-up observations were conducted for KMT-2024-BLG-0176 and KMT-2024-BLG-0349 
using telescopes from the Las Cumbres Observatory (LCO) global network.  These 
follow-up observations were carried out during the peak magnification phase of 
the microlensing events, when the lensing magnification was high.  All events 
were initially discovered by KMTNet. Following the established convention in the 
microlensing community to use the designation from the survey that first identified 
the event, we adopt the KMTNet event identifiers throughout this paper.

Images of the source stars for the events were obtained through observations 
conducted by the respective microlensing surveys. KMTNet data were collected 
using a network of three identical 1.6-meter telescopes designed for continuous 
monitoring of microlensing events. These telescopes are strategically located 
across the Southern Hemisphere: at Siding Spring Observatory in Australia (KMTA), 
Cerro Tololo Inter-American Observatory in Chile (KMTC), and the South African 
Astronomical Observatory in South Africa (KMTS). OGLE observations were carried 
out with the 1.3-meter telescope at Las Campanas Observatory in Chile, while MOA 
data were acquired using the 1.8-meter telescope at Mt. John University Observatory 
in New Zealand. The fields of view of the KMTNet, OGLE, and MOA telescopes are 
4.0, 1.4, and 2.2 square degrees, respectively.  The LCO data for KMT-2024-BLG-0176 
were obtained with a 1.0-meter telescope at the South African Astronomical Observatory 
(LCOS), and those for KMT-2024-BLG-0349 with a 1.0-meter telescope at the Siding 
Spring Observatory in Australia (LCOA).

Images were taken primarily in the Cousins $I$ band for both the KMTNet and OGLE 
surveys, while the MOA survey employed a custom broad $R$-band filter. In all 
three surveys, a subset of images was also collected in the $V$ band to enable 
measurement of the source color. LCO follow-up observations were conducted in 
the $I$ band.

The lensing light curves of the events were constructed by performing photometry 
on the source stars. The photometric measurements were carried out using codes 
based on the difference imaging technique \citep{Tomaney1996, Alard1998}, with 
survey-specific implementations.  For the KMTNet data, we used a Python-based 
image subtraction code developed by \citet{Albrow2009}. The OGLE data were 
processed with the DIA photometry pipeline developed by \citet{Wozniak2000}, 
while the MOA data were reduced using a custom difference imaging pipeline 
\citep{Bond2001}.  To ensure the best data quality, we reprocessed the KMTNet 
data using the updated photometric pipeline developed by \citet{Yang2024}. The 
LCO data were processed using the KMTNet pipeline.  The error bars of the photometry 
data derived from the different pipelines were rescaled to ensure consistency with 
the scatter in the measurements and to bring the reduced chi-square value ($\chi^2$ 
per degree of freedom) of the best-fit model for each data set close to unity. This 
error bar normalization was carried out following the procedure described in 
\citet{Yee2012}.

Short-duration features in KMTNet data were not attributed to correlated noise 
in our analysis. Such noise can arise from residual images of previous exposures, 
but microlensing light curves rarely mimic these patterns. On the uncommon 
occasions when a model matches very short-duration features suggestive of 
correlated noise, we directly inspect the corresponding images to verify the 
signal. No such spurious cases were identified for the events analyzed here.

\section{Light curve analyses} \label{sec:three}

For all events, the light curves exhibit short-duration anomalies superimposed on 
otherwise smooth and symmetric 1L1S-like profiles.  The planetary origin of these 
anomalies was identified through detailed modeling of the lensing light curves. 
In planetary lensing events, such short-term anomalies arise when the source
passes over or near a small caustic created by the planet.

Modeling the light curve of a microlensing event caused by a lens system consisting 
of two masses (binary-lens single-source, or 2L1S) requires seven fundamental 
parameters.  The first three parameters ($t_0$, $u_0$, and $\te$) characterize 
the approach of the lens and source, representing the time of closest approach, 
the impact parameter (normalized to the angular Einstein radius $\thetae$), and 
the event timescale, respectively. To model the anomaly caused by the planet, 
three additional parameters are used: $s$, $q$, and $\alpha$.  These parameters 
denote the projected separation between the planet and its host (normalized to 
$\thetae$), the mass ratio between the planet and host, and the angle between 
the source trajectory and the axis connecting the planet and host, respectively. 
When the source crosses or closely approaches a caustic, finite source effects 
become significant. To account for this, an additional parameter, $\rho$, 
representing the normalized source radius defined as the ratio of the angular 
source radius to $\thetae$, is included in the modeling.

Modeling of the events was conducted by searching for the set of lensing parameters 
that best describe the observed light curves.  For the search of the 1L1S parameters 
$(t_0, u_0, \te, \rho)$, we employed a downhill optimization method, as the lensing 
magnification varies smoothly with changes in these parameters. In contrast, for the 
binary-lens parameters $(s, q, \alpha)$, the magnification can change discontinuously 
with small variations, making the parameter space more complex to explore.  To handle 
this, we adopted a hybrid method in which the binary-lens parameters were explored 
using a grid search, while the remaining parameters were optimized using a downhill 
approach. Grid-based exploration of the binary-lens parameters also helps reveal 
degenerate solutions. In the following subsections, we present the modeling results 
for each individual event.

\begin{figure}[t]
\includegraphics[width=\columnwidth]{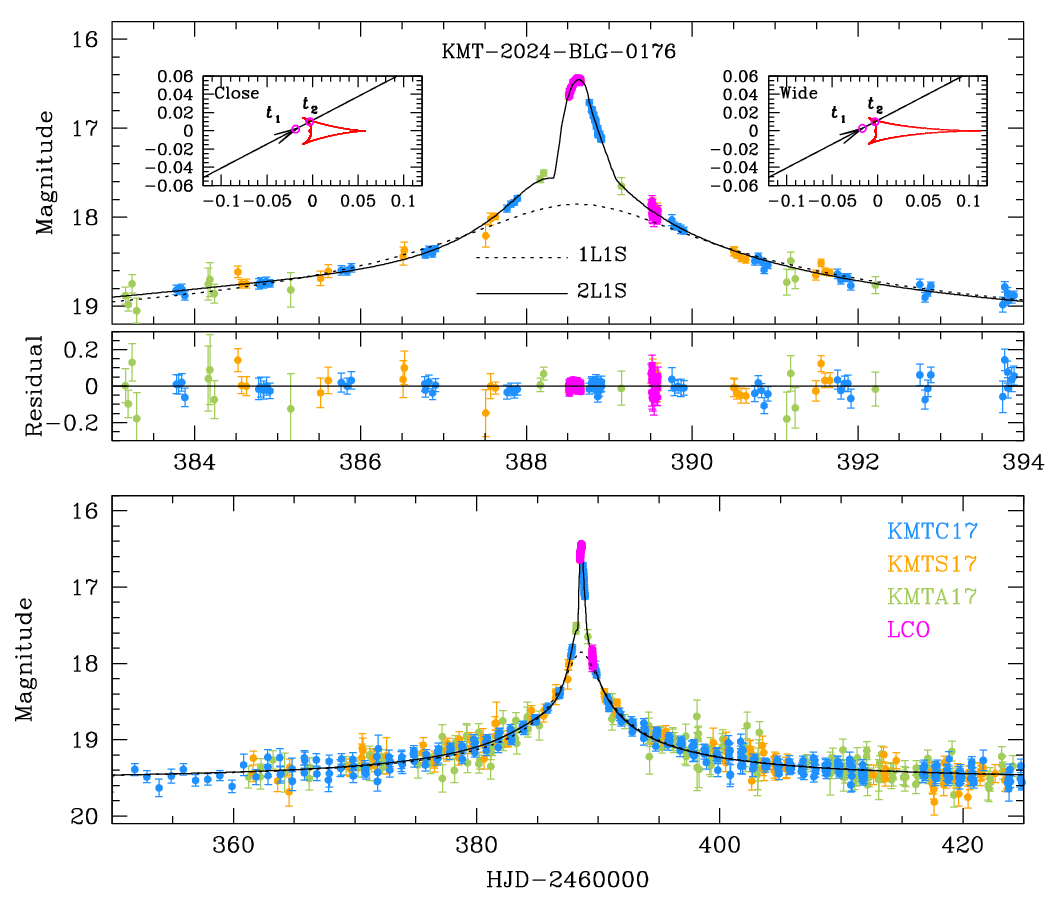}
\caption{
Lensing light curve of KMT-2024-BLG-0176.  The lower panel presents the full light 
curve of the event, while the upper panel provides a zoomed-in view around the peak. 
The solid curve corresponds to the best-fit 2L1S model (wide solution), and the 
dotted curve shows a 1L1S model derived by excluding the data affected by the anomaly. 
The two insets in the upper panel illustrate the lens system configurations for the 
close and wide solutions. In each diagram, the red cusped figure represents the 
caustic, and the arrowed line denotes the source trajectory.  Coordinates are centered 
at the position of the primary lens and lengths are scaled to the Einstein radius. 
}
\label{fig:one}
\end{figure}

\subsection{KMT-2024-BLG-0176} \label{sec:three-one}

The microlensing event KMT-2024-BLG-0176 was initially detected by the KMTNet survey. 
It involved a source star with a baseline magnitude of $I_{\rm base} = 20.1$, located 
in a region with $I$-band extinction of $A_I = 1.36$ according to the reddening map 
constructed by \citet{Gonzalez2012}.  The lensing-induced brightening was first identified 
on 2024 March 18 (HJD$^\prime \equiv {\rm HJD} - 2460000 = 387$), one day prior to the peak. 
The event reached its peak brightness, achieving a very high magnification of approximately 
$A_{\rm max} \sim 160$.  Given that the peak region of a high-magnification event is highly 
sensitive to planetary perturbations, the LCO group conducted follow-up observations on two 
consecutive nights, HJD$^\prime =388$ and 389.  The source lies within KMTNet field BLG17, 
which was observed at a cadence of one frame per hour.

The lensing light curve of the event is presented in Figure~\ref{fig:one}, with 
a close-up view of the peak region shown in the upper panel. Near the time of 
peak magnification, the light curve exhibits a clear anomaly that deviates from 
the standard 1L1S model. This anomaly, which lasts for about three days, is marked 
by positive deviations relative to the 1L1S prediction. It consists of two distinct 
features: a prominent bump centered at HJD$^\prime \sim 388.66$ ($t_2$), and a 
weaker bump occurring shortly before the main feature (HJD$^\prime \sim 387.80$, 
$t_1$).

Central anomalies in high-magnification microlensing events can arise from either 
a planetary companion located near the Einstein ring of the primary lens \citep{Griest1998} 
or a binary companion with a very close or wide separation \citep{Han2009}. To determine 
the origin of the anomaly observed in this event, we modeled the light curve under 
a 2L1S configuration. In this analysis, we explored a wide range of parameter space 
for the binary separation and mass ratio, specifically covering $-1.0 < \log s < 1.0$ 
and $-5 < \log q < 1.0$, to test both the planetary and binary companion scenarios.

\begin{table}[t]
\caption{Lensing parameters of KMT-2024-BLG-0176.\label{table:two}}
\begin{tabular*}{\columnwidth}{@{\extracolsep{\fill}}lllll}
\hline\hline
\multicolumn{1}{c}{Parameter}        &
\multicolumn{1}{c}{Close}            &
\multicolumn{1}{c}{Wide}             \\
\hline
 $\chi^2$               &  $1659.4             $  &  $1653.8               $   \\  
 $t_0$ (HJD$^\prime$)   &  $388.5232 \pm 0.0090$  &  $388.52034 \pm 0.01075$   \\
 $u_0$                  &  $0.0102 \pm 0.0012  $  &  $0.0102 \pm 0.0013    $   \\
 $\te$ (days)           &  $53.05 \pm 4.41     $  &  $55.24 \pm 5.79       $   \\
 $s$                    &  $0.7534 \pm 0.0084  $  &  $1.3229 \pm 0.0201    $   \\
 $q$  (10$^{-3}$)       &  $6.98 \pm 0.89      $  &  $6.96 \pm 0.94        $   \\
 $\alpha$ (rad)         &  $2.659 \pm 0.015    $  &  $2.660 \pm 0.018      $   \\
 $\rho$ (10$^{-3}$)     &  $3.92 \pm 0.39      $  &  $3.77 \pm 0.42        $   \\
\hline             
\end{tabular*}
\tablecomments{HJD$^\prime = {\rm HJD} - 2460000$.}
\end{table}

A detailed analysis of the light curve indicates that the anomaly is of planetary 
origin. The modeling yields a pair of degenerate solutions resulting from the 
well-known ``close–wide degeneracy'' \citep{Griest1998}. The lensing parameters 
for both the close and wide solutions, along with their corresponding $\chi^2$ 
values, are listed in Table~\ref{table:two}. The degeneracy is strong, with the 
wide solution favored only marginally over the close solution by $\Delta\chi^2 = 5.6$. 
In Figure~\ref{fig:one}, the model curve corresponding to the wide solution is plotted 
over the data points.

The best-fit binary-lens parameters are $(s, q) \sim (0.75, 7.0 \times 10^{-3})$ for 
the close solution and $(s, q) \sim (1.32, 7.0 \times 10^{-3})$ for the wide solution.  
The binary separations of the two solutions approximately follow the relation $s_{\rm c} 
= 1/s_{\rm w}$, where $s_{\rm c}$ and $s_{\rm w}$ denote the separations of the close 
and wide solutions, respectively. The inferred mass ratios are roughly seven times the 
Jupiter-to-Sun mass ratio, indicating that the lower-mass component of the lens system 
is a giant planet. The event timescale is measured to be $\te \sim 54$ days.  As 
discussed below, the main anomaly was caused by the source crossing a caustic induced 
by the planetary companion. The normalized source radius was determined from modeling 
this feature.

The two insets in the upper panel of Figure~\ref{fig:one} show the lens-system 
configurations for the close and wide solutions, illustrating the source trajectory 
relative to the caustic. In both cases, the planet produces a similar central caustic 
structure with four cusps.  The source first approached the weak cusp located along 
the star-planet axis, then crossed the caustic near the tip of its upper-left section.  
This caustic crossing produced the main bump (at around $t_2$) in the light curve, 
while the earlier approach to the weak cusp caused the smaller bump (at around $t_1$) 
preceding the main anomaly.  We mark the source positions at times $t_1$ and $t_2$ 
with empty circles, whose sizes are scaled to represent the source size.

\begin{figure}[t]
\includegraphics[width=\columnwidth]{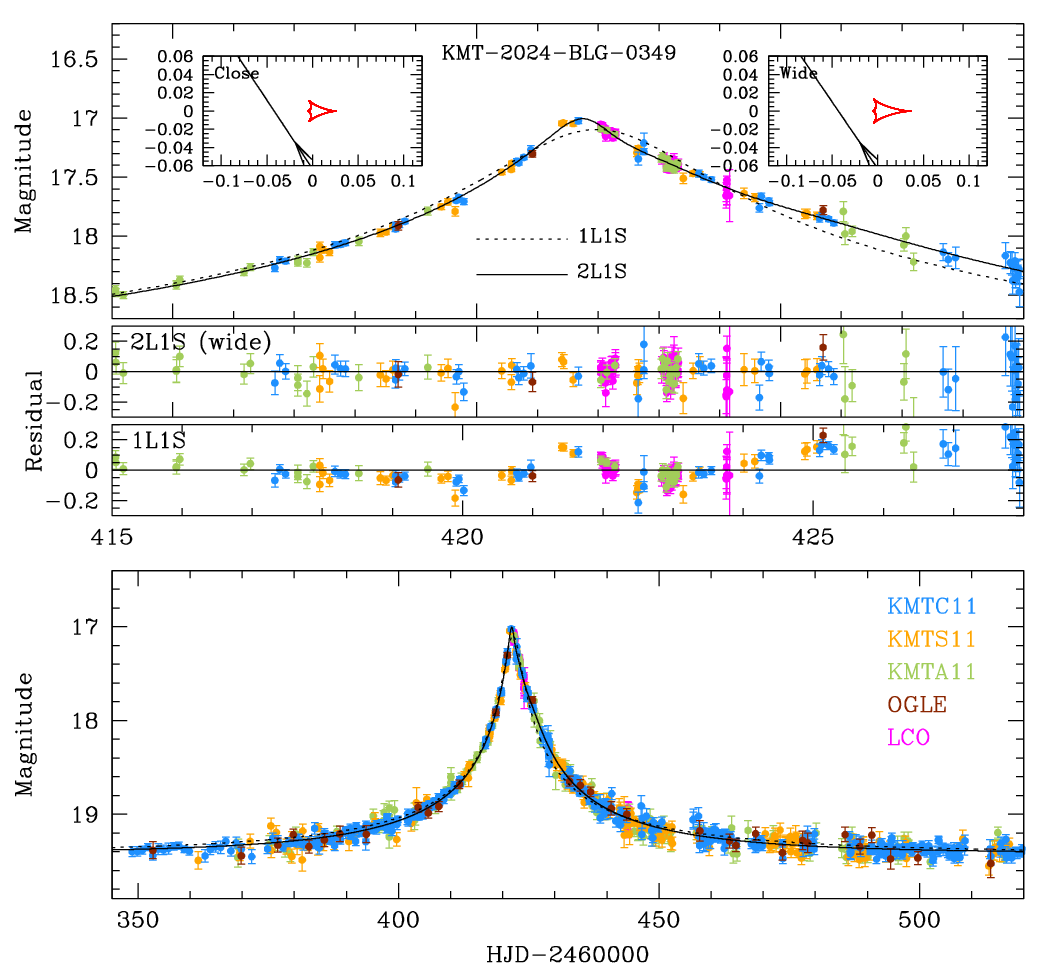}
\caption{
Light curve of KMT-2024-BLG-0349.
The notations follow those used in Fig.~\ref{fig:one}.
}
\label{fig:two}
\end{figure}

\subsection{KMT-2024-BLG-0349} \label{sec:three-two}

The source of the microlensing event KMT-2024-BLG-0349 had a baseline magnitude
of $I_{\rm base} = 19.41$ prior to lensing magnification. The event was initially 
detected by the KMTNet survey on 2024 March 29 (HJD$^\prime = 398$) and was 
subsequently confirmed by the OGLE group.  The peak region of the light curve was 
continuously monitored by the LCO follow-up group over three consecutive nights 
(HJD$^\prime =$  422, 423, and 424), with an additional observation conducted on 
HJD$^\prime =$ 444.  The $I$-band extinction toward the event location is $A_I = 
2.64$.  The event reached its peak on HJD$^\prime = 422$ (2024 April 21) with a 
moderate maximum magnification of $A_{\rm max} \sim 30$.  The source lies in the 
KMTNet BLG11 field, which was monitored with a cadence of 2.5 hours.

Figure~\ref{fig:two} presents the full light curve (bottom panel) and a close-up 
view of the peak region (upper panel) for the lensing event. At first glance, the 
light curve appears consistent with a 1L1S event, exhibiting a smooth brightness 
profile. However, closer examination reveals that the 1L1S model does not fully 
reproduce the observed data, leaving residuals near the peak. In particular, the 
peak region displays a clear asymmetry relative to the time of maximum magnification.

\begin{table}[t]
\caption{Lensing parameters of KMT-2024-BLG-0349.\label{table:three}}
\begin{tabular*}{\columnwidth}{@{\extracolsep{\fill}}lllll}
\hline\hline
\multicolumn{1}{c}{Parameter}        &
\multicolumn{1}{c}{Close}            &
\multicolumn{1}{c}{Wide}             \\
\hline
 $\chi^2$               &  $1243.5           $  &  $1241.2           $  \\  
 $t_0$ (HJD$^\prime$)   &  $422.325 \pm 0.020$  &  $422.340 \pm0.022 $  \\
 $u_0$                  &  $0.0350 \pm 0.0026$  &  $0.0357 \pm 0.0027$  \\
 $\te$ (days)           &  $52.29 \pm 3.11   $  &  $52.67 \pm 3.02   $  \\
 $s$                    &  $0.528 \pm 0.009  $  &  $1.847 \pm 0.040  $  \\
 $q$  (10$^{-3}$)       &  $16.54 \pm 2.01   $  &  $17.93 \pm 2.52   $  \\
 $\alpha$ (rad)         &  $0.985 \pm 0.020  $  &  $0.972 \pm 0.025  $  \\
 $\rho$ (10$^{-3}$)     &  $< 12             $  &  $< 12             $  \\
\hline             
\end{tabular*}
\end{table}

Because deviations near the peak of a high-magnification event can be indicative 
of a planetary companion, we performed a detailed analysis under a 2L1S framework. 
This modeling yielded two planetary solutions with parameters 
$(s, q) \sim (0.53, 16.5 \times 10^{-3})$ for the close model and 
$(s, q) \sim (1.85, 17.9 \times 10^{-3})$ for the wide model, confirming that 
the observed anomaly was indeed caused by a companion with a planetary mass. 
The complete sets of lensing parameters for both solutions are listed in Table 
\ref{table:three}.  For this event, the normalized source radius could not be 
precisely measured because the anomaly did not result from caustic crossings, and 
only an upper limit could be derived.  The two solutions are strongly degenerate, 
with the wide solution preferred only slightly, by $\Delta\chi^2 = 2.3$.  In Figure
\ref{fig:two}, the model curve corresponding to the wide solution is plotted 
over the data. The planet-host separations approximately satisfy the relation 
$s_{\rm c} \sim 1/s_{\rm w}$, indicating that the degeneracy arises from the 
well-known close-wide ambiguity.

The lens-system configurations for the close and wide solutions are shown in the 
insets of the upper panel of Figure~\ref{fig:two}. In both cases, the planet 
produces a small central caustic with four cusps. The observed anomaly arises as 
the source passes behind the caustic. The peak of the light curve corresponds to 
the source traversing a region of positive deviation extending from the lower cusp 
of the caustic, just before its closest approach to the host.  This is followed by 
the source moving through an extended region of negative deviation. The combined 
effect of these features results in the asymmetry observed near the peak of the 
light curve.

\subsection{KMT-2024-BLG-1870} \label{sec:three-three}

The microlensing event KMT-2024-BLG-1870 was observed by all three microlensing 
surveys currently in operation. It was initially detected by the KMTNet survey 
during its early magnification phase on 2024 July 17 (HJD$^\prime = 508$), and 
was later independently identified by both the OGLE and MOA surveys. The event 
occurred within a narrow overlapping region of the KMTNet BLG31 and BLG32 fields, 
each monitored at a cadence of one observation every 2.5 hours. The source star is 
relatively bright, with a baseline magnitude of $I_{\rm base} = 16.97$, and the 
$I$-band extinction toward the event location is $A_I = 1.48$.

\begin{figure}[t]
\includegraphics[width=\columnwidth]{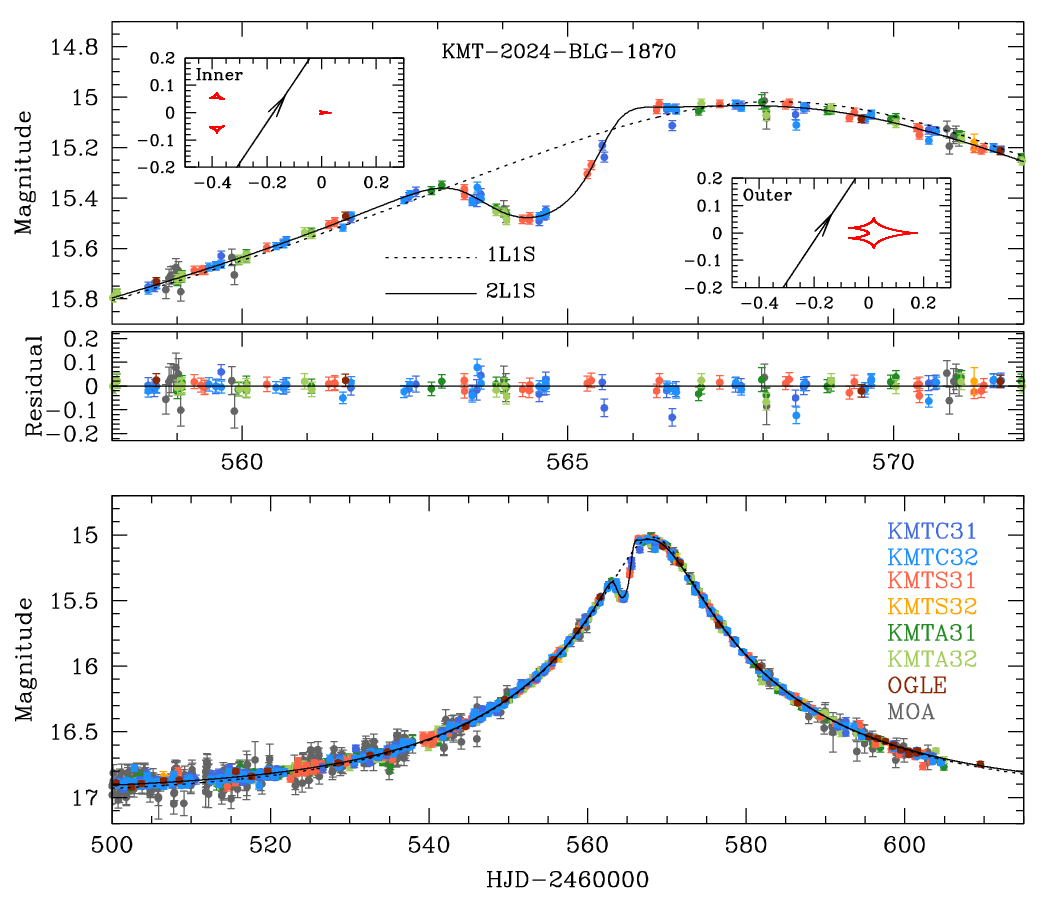}
\caption{
Light curve of KMT-2024-BLG-1870.
}
\label{fig:three}
\end{figure}

Figure~\ref{fig:three} presents the lensing light curve of the event, which ends 
at HJD$^\prime = 609$, corresponding to the conclusion of the 2024 bulge season. 
The light curve displays an anomaly occurring four days prior to the peak, marked 
by a brief dip feature centered at HJD$^\prime \sim 564.5$. This feature strongly 
suggests the presence of a planetary companion to the lens, similar to those 
observed in events such as MOA-2022-BLG-033, KMT-2023-BLG-0119, and KMT-2023-BLG-1896 
\citep{Han2025a}.

Detailed modeling of the light curve confirmed that the anomaly was caused by a 
planetary companion. The analysis yielded a pair of degenerate solutions resulting 
from the so-called ``inner-outer degeneracy,'' which occurs when the source 
trajectory passes on either the inner or outer side of a peripheral caustic 
induced by the planet \citep{Gaudi1997, Herrera2020, Yee2021, Zhang2022}.  The 
lensing parameters corresponding to the two solutions are listed in 
Table~\ref{table:four}. The outer solution is preferred over the inner solution 
by $\Delta\chi^2 = 22.0$, indicating that the degeneracy is relatively mild.

\begin{table}[t]
\caption{Lensing parameters of KMT-2024-BLG-1870.\label{table:four}}
\begin{tabular*}{\columnwidth}{@{\extracolsep{\fill}}lllll}
\hline\hline
\multicolumn{1}{c}{Parameter}        &
\multicolumn{1}{c}{Inner}            &
\multicolumn{1}{c}{Outer}            \\
\hline
 $\chi^2$               &  $2459.1           $  &  $2437.1           $  \\  
 $t_0$ (HJD$^\prime$)   &  $568.232 \pm 0.013$  &  $568.194 \pm 0.013$  \\
 $u_0$                  &  $0.1463 \pm 0.0019$  &  $0.1503 \pm 0.0020$  \\
 $\te$ (days)           &  $37.20 \pm 0.33   $  &  $36.66 \pm 0.34   $  \\
 $s$                    &  $0.8259 \pm 0.0036$  &  $1.0089 \pm 0.0042$  \\
 $q$  (10$^{-3}$)       &  $1.515 \pm 0.070  $  &  $1.470 \pm 0.059  $  \\
 $\alpha$ (rad)         &  $2.1608 \pm 0.0036$  &  $2.1564 \pm 0.0034$  \\
 $\rho$ (10$^{-3}$)     &  $< 15             $  &  $< 15             $  \\
\hline             
\end{tabular*}
\end{table}

For solutions affected by the inner-outer degeneracy, the planet-host separations 
of the inner ($s_{\rm in}$) and outer ($s_{\rm out}$) solutions satisfy the 
relation derived by \citet{Hwang2022} and \citet{Gould2022}, given by
\begin{equation}
\sqrt{s_{\rm in}\times s_{\rm out} } = s^\dagger; \qquad
s^\dagger  = { \sqrt{u_{\rm anom}^2+4} \pm u_{\rm anom}\over 2}.
\label{eq1}
\end{equation}
\hskip-3pt
Here, $u_{\rm anom} = (\tau^2_{\rm anom} + u_0^2)^{1/2}$, $\tau_{\rm anom} = 
(t_{\rm anom} - t_0)/\te$, and $t_{\rm anom}$ denotes the time of the anomaly. 
The sign in the last term is “$+$” for anomalies exhibiting a bump feature (positive 
anomaly) and “$-$” for those with a dip feature. Using the parameters $(t_0, u_0, \te) 
\sim (568.232, 0.1463, 37.20)$, we obtain $s^\dagger = 0.914$, which closely matches 
the geometric mean $\sqrt{s_{\rm in} \times s_{\rm out}} = 0.912$. This agreement 
confirms that the similarity between the two solutions arises from the inner-outer 
degeneracy. The normalized source radius could not be measured, and only an upper 
limit could be placed.  \citet{Hwang2022} presented an analytic expression for 
heuristically estimating the mass ratio of a planetary companion that induces a 
dip feature in the light curve: 
\begin{equation} 
q = \left(
{\Delta t_{\rm dip} \over 4\te}
\right)^2\left( {\sin^2\alpha \over u_{\rm anom}}\right),
\label{eq2}
\end{equation}
\hskip-3pt
where $\Delta t_{\rm dip}$ represents the duration of the dip anomaly.  Applying 
this relation with $\Delta t_{\rm dip} \sim 2.7$ days yields a heuristic estimate 
of the planet-to-host mass ratio, $q \simeq 1.6 \times 10^{-3}$, which is in good 
agreement with the value derived from the detailed modeling.

The two insets in the upper panel of Figure~\ref{fig:three} illustrate the lens-system 
configurations for the inner and outer solutions. In the inner solution, the central 
and peripheral caustics are clearly separated, whereas in the outer solution, the two 
caustics merge to form a single resonant caustic.  The anomaly in the inner solution 
is caused by the source passing on the inner side of the peripheral caustic, while 
in the outer solution, it results from the source passing on the outer side.

\subsection{KMT-2024-BLG-2087} \label{sec:three-four}

The microlensing event KMT-2024-BLG-2087 was initially discovered by the KMTNet 
survey. Although it was not originally identified by MOA, it was later determined 
that the source lay within the footprint of a MOA survey field. Using postseason 
photometry, we were able to recover data for the event from MOA images.  The source 
star has a baseline magnitude of $I_{\rm base} = 18.4$, and the $I$-band extinction 
in its direction is $A_I = 1.89$. The event occurred in the overlapping region of 
the KMTNet prime fields BLG02 and BLG42, which were monitored with a combined cadence 
of 0.25 hours, enabling dense temporal coverage. The onset of lensing-induced brightening
 was detected at an early stage on 2024 August 5 (HJD$^\prime = 508$).

\begin{figure}[t]
\includegraphics[width=\columnwidth]{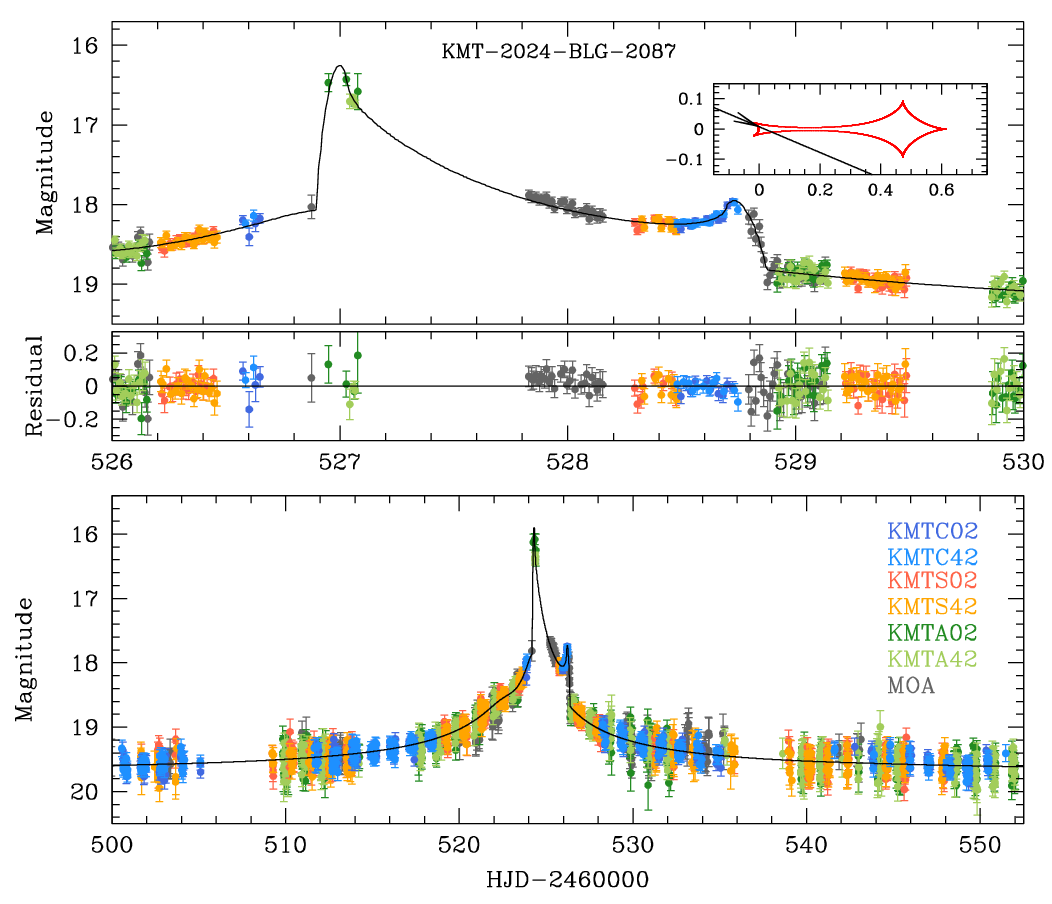}
\caption{
Light curve of the lensing event KMT-2024-BLG-2087.
}
\label{fig:four}
\end{figure}

\begin{table}[t]
\caption{Lensing parameters of KMT-2024-BLG-2087.\label{table:five}}
\begin{tabular*}{\columnwidth}{@{\extracolsep{\fill}}lllll}
\hline\hline
\multicolumn{1}{c}{Parameter}        &
\multicolumn{1}{c}{Value}             \\
\hline
 $\chi^2$               &  $3879.4             $  \\  
 $t_0$ (HJD$^\prime$)   &  $527.220  \pm 0.009 $  \\
 $u_0$                  &  $0.00674 \pm 0.00065$  \\
 $\te$ (days)           &  $36.57 \pm 3.98     $  \\
 $s$                    &  $1.2623 \pm 0.0090  $  \\
 $q$  (10$^{-3}$)       &  $8.80 \pm 1.46      $  \\
 $\alpha$ (rad)         &  $3.542 \pm 0.015    $  \\
 $\rho$ (10$^{-3}$)     &  $1.20 \pm 0.16      $  \\
\hline             
\end{tabular*}
\end{table}

\begin{table*}[t]
\centering
\caption{Source colors and magnitudes, angular Einstein radii, and relative proper motions.  \label{table:six}}
\begin{tabular}{llllllllllcc}
\hline\hline
\multicolumn{1}{c}{Parameter}             &
\multicolumn{1}{c}{KMT-2024-BLG-0176}     &
\multicolumn{1}{c}{KMT-2024-BLG-0349}     &
\multicolumn{1}{c}{KMT-2024-BLG-1870}     &
\multicolumn{1}{c}{KMT-2024-BLG-2087}     \\
\hline
 $(V-I)_{\rm S}$           &  $2.262 \pm 0.015 $    &   $3.068 \pm 0.091 $   &  $2.249 \pm 0.017 $     &  $2.873 \pm 0.183 $   \\
 $I_{\rm S}$               &  $22.010 \pm 0.004$    &   $21.481 \pm 0.008$   &  $17.223 \pm 0.002$     &  $22.292 \pm 0.017$   \\
 $(V-I, I)_{\rm RGC}$      &  $(2.314, 16.320) $    &   $(3.389, 17.191) $   &  $(2.255, 16.092) $     &  $(2.782, 16.495) $   \\
 $(V-I, I)_{{\rm RGC},0}$  &  $(1.060, 14.339) $    &   $(1.060, 14.596) $   &  $(1.060, 14.317) $     &  $(1.060, 14.413) $   \\
 $(V-I)_{{\rm S},0}$       &  $1.008 \pm 0.043 $    &   $0.739 \pm 0.099 $   &  $1.054 \pm 0.043 $     &  $1.151 \pm 0.183 $   \\
 $I_{{\rm S},0}$           &  $20.029 \pm 0.020$    &   $18.887 \pm 0.022$   &  $15.448 \pm 0.020$     &  $20.210 \pm 0.017$   \\
 Spectral type             &   K3V                  &    G5V                 &   K3III                 &   K6.5V               \\
 $\theta_*$ ($\mu$as)      &  $0.435 \pm 0.036 $    &   $0.542 \pm 0.066 $   &  $3.85 \pm 0.32   $     &  $0.467 \pm 0.091 $   \\
 $\thetae$ (mas)           &  $0.106 \pm 0.015 $    &   $> 0.05          $   &  $> 0.26          $     &  $0.296 \pm 0.074 $   \\
 $\mu$ (mas/yr)            &  $0.76 \pm 0.11   $    &   $> 1.0           $   &  $> 2.6           $     &  $3.417 \pm 0.923 $   \\
\hline                                                             
\end{tabular}                                                      
\end{table*}

\begin{figure*}[t]
\centering
\includegraphics[width=15.0cm]{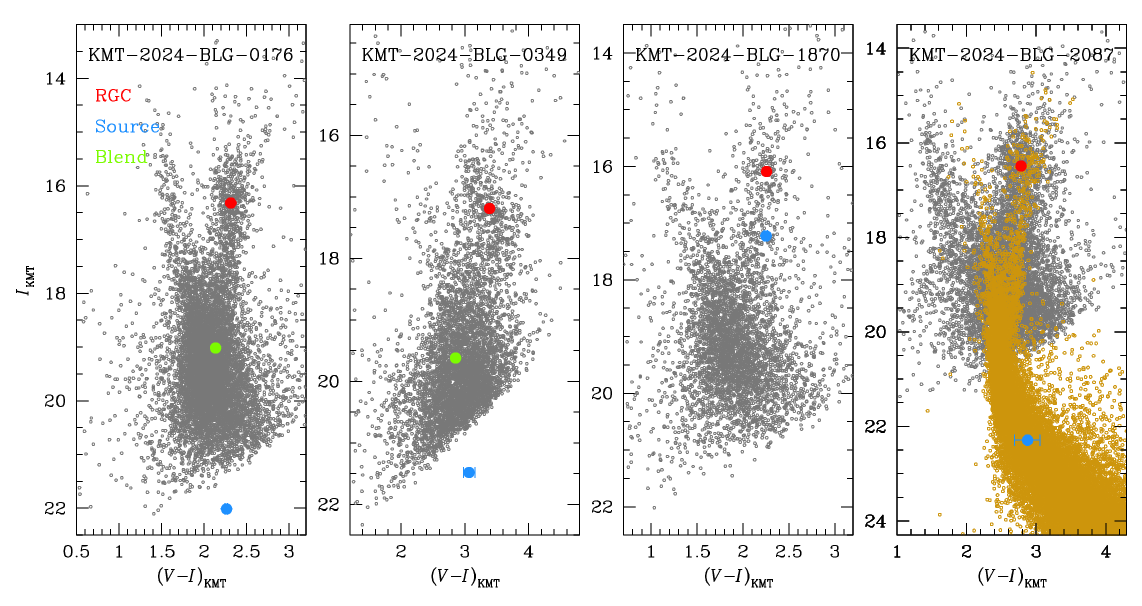}
\caption{
Locations of source and RGC centroid in the instrumental color-magnitude diagrams. 
The position of the blend is also indicated in green for events where the blend flux was 
measured.
}
\label{fig:five}
\end{figure*}

The light curve of the event is presented in Figure~\ref{fig:four}. Near the peak, the 
light curve deviates from the standard 1L1S shape, displaying an anomaly marked by two 
abrupt features occurring around HJD$^\prime = 527.0$ and 528.7. Although the coverage 
of this anomalous region is incomplete due to unfavorable weather conditions at both 
the KMTC and KMTS sites, the sharp changes in magnification at these two epochs, along 
with the intervening ``U''-shape light curve profile, indicate that the anomaly was likely 
caused by caustic crossings of the source.

Given the caustic-induced features in the light curve, we modeled the event using a 
2L1S configuration. The analysis yielded a unique solution with binary parameters of 
$(s, q) \sim (1.26, 8.8 \times 10^{-3})$.  An examination of the local close solution 
with $s < 1$ showed that the wide solution is significantly favored, with a $\Delta\chi^2 
= 34.3$ compared to the close solution.  The mass ratio is approximately ten times larger 
than that of the Jupiter-Sun system. The event timescale is estimated to be approximately  
$\te \sim 37$ days.  Since microlensing events with comparable timescales are typically 
caused by low-mass stellar lenses \citep{Han2003}, and considering that the upper bound 
for planetary mass is around $13M_{\rm J}$, the companion is likely to be a giant planet.  
The complete set of lensing parameters is provided in Table~\ref{table:five}. While the 
uncertainty in the measurement is relatively large, the normalized source radius is 
nonetheless reasonably constrained.  In addition to the caustic-crossing features, the 
light curve shows a subtle bump centered around HJD$^\prime = 522$.

The lens-system configuration is shown in the inset of the upper panel of Figure 
\ref{fig:four}.  A planet located near the Einstein ring produces a single resonant 
caustic, in which the central and peripheral caustics are connected by a narrow bridge.  
The source made a diagonal passage through the central caustic, entering at the upper-left 
fold and exiting at the lower fold, giving rise to the sharp features observed in the 
anomaly.  Prior to the caustic entry, the source approached the upper-left cusp closely, 
resulting in a weak bump feature that preceded the first caustic spike.

\section{Source stars and Einstein radii }  \label{sec:four}

In this section, we present the properties of the source stars associated with the 
analyzed events. Characterizing the source is crucial for a thorough understanding 
of the event. Furthermore, it allows us to estimate the angular Einstein radius using 
the relation
\begin{equation}
\thetae = \frac{\theta_*}{\rho}.
\label{eq3}
\end{equation}
\hskip-3pt
Here, $\theta_*$ denotes the angular radius of the source, which is derived from the 
color and apparent magnitude, while $\rho$ is the normalized source radius derived 
from light-curve modeling.

The source star for each microlensing event was characterized by determining its 
de-reddened color and magnitude, $(V-I, I)_{{\rm S},0}$, corrected for interstellar 
extinction and reddening. This process involved two steps. First, we derived the 
instrumental color and magnitude, $(V-I, I)_{\rm S}$, by performing a regression 
of the observed $V$- and $I$-band data against the best-fit model light curve. 
For this analysis, we utilized photometry data obtained using the pyDIA pipeline 
\citep{Albrow2017}. Second, we estimated the de-reddened color and magnitude of the 
source by referencing the centroid of the red giant clump (RGC) in the color-magnitude 
diagram (CMD). The intrinsic color and magnitude of the RGC, $(V-I, I)_{{\rm RGC},0}$, 
are well established from previous works by \citet{Bensby2013} and \citet{Nataf2013}, 
and were used as a standard for this calibration.

In the case of KMT-2024-BLG-2087, the available $V$-band data lacked the precision 
needed to provide a reliable color estimate.  For this event, we employed an 
alternative approach by combining the CMD constructed from KMTC images with that of 
stars in Baade’s Window observed by the Hubble Space Telescope \citep{Holtzman1998}. 
We then adopted the mean color of stars lying along either the giant or main-sequence 
branch within the brightness range defined by the source’s measured $I$-band magnitude 
in the combined CMD.

In Figure~\ref{fig:five}, we mark the position of each source star relative to the 
RGC centroid in the instrumental CMD.  The determined values of $(V-I, I)_{\rm S}$, 
$(V-I, I)_{\rm RGC}$, $(V-I, I)_{{\rm RGC},0}$, and $(V-I, I)_{{\rm S},0}$ are 
presented in Table~\ref{table:six}.  The 
source of KMT-2024-BLG-1870 is identified as a giant star, whereas the sources of 
the other events are main-sequence stars.

\begin{figure}[t]
\includegraphics[width=\columnwidth]{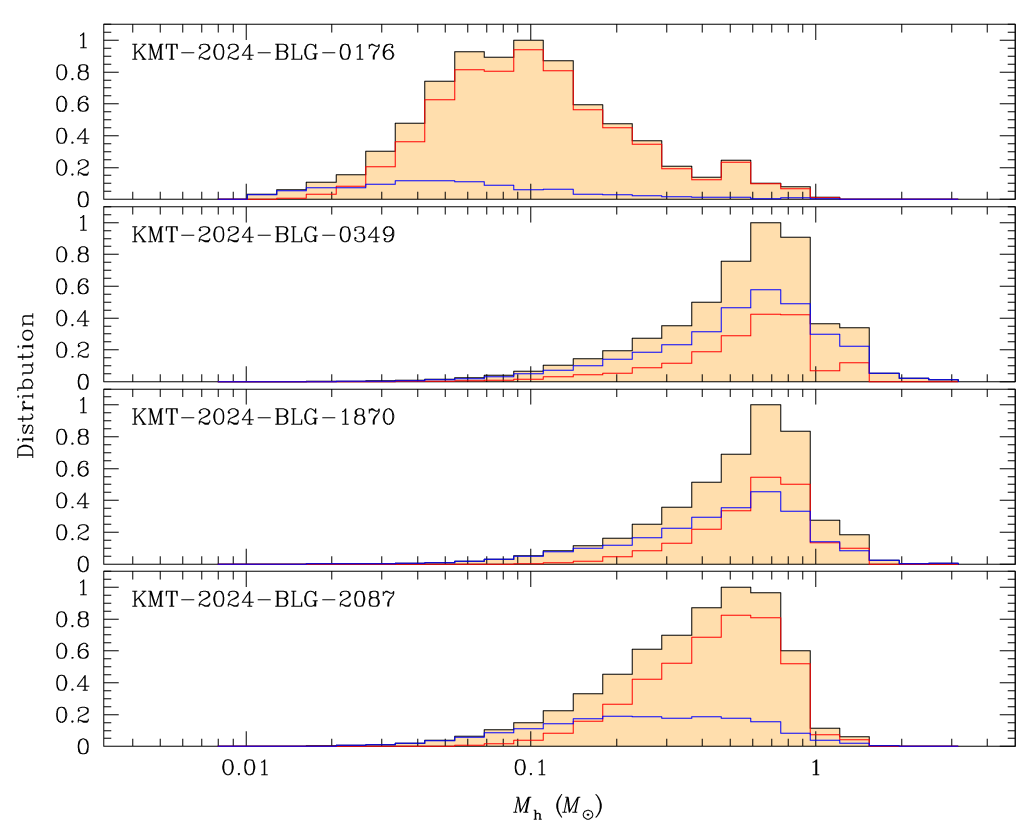}
\caption{
Bayesian Posteriors of lens mass.  In each panel, the blue and red curves represent 
the contributions from disk and bulge lenses, respectively, while the black curve 
shows the sum of both contributions.
}
\label{fig:six}
\end{figure}

We estimated the angular source radius from the dereddened color and magnitude using 
the $(V-K, V)$--$\theta_*$ relation established by \citet{Kervella2004}. To apply this 
relation, we first converted the measured $V-I$ color into $V-K$ using the color-color 
relation of \citet{Bessell1988}.\footnote{
We investigated the scatter in the source-radius estimate introduced by this color 
conversion by directly deriving $\theta_*$ from the $(V-I, I)$--$\theta_*$ relation 
provided by \citet{Adams2018}. We found that the angular source radii derived from the 
two methods agree to within 5\%. This discrepancy is smaller than the 7\% uncertainty 
adopted to account for the combined effects of differential extinction, CMD calibration, 
and the intrinsic uncertainties in the color--surface-brightness relation. }
With the estimated angular source radius, we then 
determine the angular Einstein radius using the relation given in Eq.~(\ref{eq3}). In 
Table~\ref{table:six}, we present the determined values of $\theta_*$ and $\thetae$. 
The table also lists the relative lens-source proper motion derived by $\mu = \thetae/ 
\te$.  In events for which only an upper limit on $\rho$ is constrained, we report the 
lower bounds of $\theta_{\rm E}$ and $\mu$.

\begin{table*}[t]
\centering
\caption{Physical lens parameters.  \label{table:seven}}
\begin{tabular}{llllllllllcc}
\hline\hline
\multicolumn{1}{c}{Event}                      &
\multicolumn{1}{c}{$M_{\rm h}$ ($M_\odot$)}    &
\multicolumn{1}{c}{$M_{\rm p}$ ($M_{\rm J}$)}  &
\multicolumn{1}{c}{$\dl$ (kpc)}                &
\multicolumn{2}{c}{$a_\perp$ (au)}             &
\multicolumn{1}{c}{$p_{\rm disk}$}             &
\multicolumn{1}{c}{$p_{\rm bulge}$}            \\
\hline             
KMT-2024-BLG-0176  &  $0.103^{+0.141}_{-0.054}$  &  $0.76^{+1.03}_{-0.39} $   &  $8.34^{+1.01}_{-1.13}$  &  $0.75^{+0.09}_{-0.10}$ (close)  &  $1.32^{+0.16}_{-0.18}$  (wide)   &  13\%   &  87\%    \\  [0.6ex]
KMT-2024-BLG-0349  &  $0.68^{+0.38}_{-0.38}   $  &  $11.86^{+6.56}_{-6.54}$   &  $5.75^{+2.02}_{-2.39}$  &  $1.85^{+0.65}_{-0.77}$ (close)  &  $6.46^{+2.27}_{-2.69} $ (wide)   &  64\%   &  36\%    \\  [0.6ex]
KMT-2024-BLG-1870  &  $0.67^{+0.33}_{-0.34}   $  &  $1.06^{+0.53}_{-0.55} $   &  $5.48^{+1.37}_{-2.06}$  &  $2.57^{+0.64}_{-0.97}$ (inner)  &  $3.15^{+0.79}_{-1.19}$ (outer)   &  54\%   &  46\%    \\  [0.6ex]
KMT-2024-BLG-2087  &  $0.48^{+0.27}_{-0.33}   $  &  $4.39^{+3.07}_{-2.46} $   &  $7.19^{+0.95}_{-1.28}$  &  $3.03^{+0.42}_{-0.54}$          &                                   &  29\%   &  71\%    \\  [0.6ex]
\hline              
\end{tabular}
\end{table*}

\section{Physical parameters} \label{sec:five} 

In this section, we estimate the physical parameters of the lens system, including its mass 
and distance. These physical parameters can be constrained using the lensing observables: 
the event timescale, angular Einstein radius, and microlens parallax ($\pie$).  These 
observables are related to the lens mass ($M$) and distance ($\dl$) through the relation
\begin{equation}
\te = {\thetae \over \mu}; \qquad
\thetae = (\kappa M \pi_{\rm rel})^{1/2}; \qquad
\pie = {\pi_{\rm rel} \over \thetae}.
\label{eq4}
\end{equation}
\hskip-5pt
Here, $\kappa = 4G/(c^2 {\rm au})\simeq 8,14~{\rm mas}/M_\odot$ and $\pi_{\rm rel} = 
{\rm au}(\pi_{\rm L} - \pi_{\rm S})$, where $\pi_{\rm L} = {\rm au}/D_{\rm L}$ and 
$\pi_{\rm S} = {\rm au}/D_{\rm S}$ represent the parallaxes of the lens and source, 
respectively. When all three observables are measured, the lens mass and distance can 
be uniquely determined using the relation of \citet{Gould2000}:
\begin{equation}
M = {\thetae \over \kappa\pie};\qquad
\dl = {{\rm au} \over \pie\thetae + \pi_{\rm S}}.
\label{eq5}
\end{equation}

Among the three lensing observables, the event timescale is measured for all events. 
The angular Einstein radius is determined for KMT-2024-BLG-0176, and KMT-2024-BLG-2087, 
while for the other events, it is either unconstrained or only lower limits are available.  
In contrast, the microlens parallax is not measured for any of the events, either because 
the event durations are not long enough or because the photometric precision is insufficient 
to detect the subtle deviations induced by this higher-order effect.  Given the incomplete 
set of lensing observables, we constrained the physical parameters through Bayesian analyses.

\begin{figure}[t]
\includegraphics[width=\columnwidth]{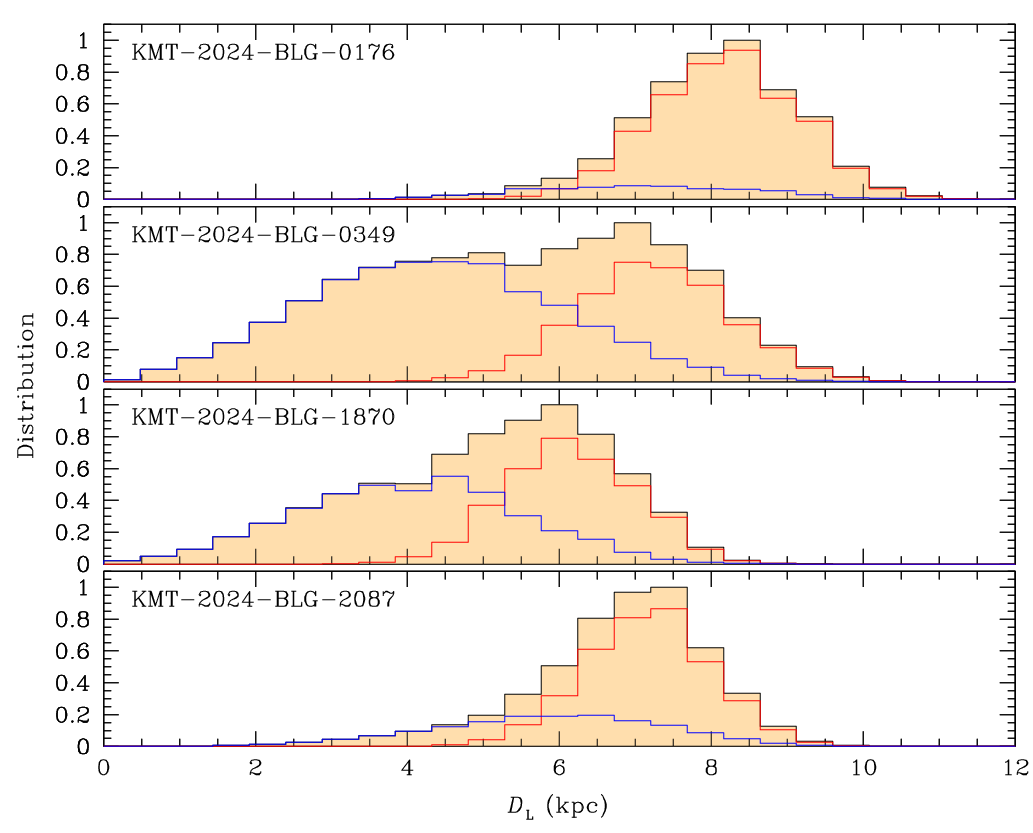}
\caption{
Bayesian Posteriors of distances to the planetary systems.
}
\label{fig:seven}
\end{figure}

The Bayesian analysis was conducted according to the following procedure. In the first 
step, we generated a large ensemble of artificial microlensing events through a Monte 
Carlo simulation. For each simulated event, the physical parameters $(M_i, D_{{\rm L},i}, 
D_{{\rm S},i}, \mu_i)$ were assigned based on a prior Galactic model and a mass function. 
The Galactic model describes the spatial and kinematic distributions of Galactic objects, 
while the mass function specified the distribution of their masses.   
For our analysis, we adopted the mass function from \citet{Jung2022} and the Galactic 
model from \citet{Jung2021}. The mass function combines the bulge initial mass function 
with the present-day disk mass function of \citet{Chabrier2003}, and includes remnants 
(white dwarfs, neutron stars, black holes) following \citet{Gould2000}. The Galactic 
model consists of a triaxial bar-shaped bulge from \citet{Han2003}, constrained by 
infrared star counts, and a modified double-exponential disk from \citet{Bennett2014}. 
Bulge kinematics are based on Gaia data \citep{Gaia2016, Gaia2018}, while disk 
kinematics follow the Gaussian prescription of \citet{Han1995}, adjusted to match the 
\citet{Bennett2014} disk.

In the second step, we computed the lensing observables, the event timescale ($t_{{\rm E},i}$) 
and angular Einstein radius ($\theta_{{\rm E},i}$), corresponding to the physical lens 
parameters using the relations given in Eq.~(\ref{eq4}). In the last step, we constructed 
posteriors of the lens mass and distance. This was achieved by assigning a weight to each 
artificial events of
\begin{equation}
w_i = \exp \left(- {\chi_i^2 \over 2}\right);\ \
\chi_i^2 = 
\left[ {t_{{\rm E},i}-\te  \over \sigma(\te)} \right]^2 +
\left[ {\theta_{{\rm E},i}-\thetae  \over \sigma(\thetae)} \right]^2,
\label{eq6}
\end{equation}
\hskip-4pt
where $(t_{\rm E}, \theta_{\rm E})$ are the measured values of the event timescale and 
angular Einstein radius, and $\sigma(t_{\rm E})$, $\sigma(\theta_{\rm E})$ are their 
respective uncertainties.
In computing the weight, we assume a weak correlation between $\te$ and $\thetae$. 
This is reasonable because $\te$ is determined directly from light-curve modeling, 
while $\thetae$ is obtained independently from the source's angular radius 
$\theta_*$, inferred from its color and brightness. Although $\thetae = \theta_*/
\rho$ introduces a possible correlation through $\rho$, in practice the dependence 
is weak since $\thetae$ relies primarily on independent source property estimates.

The posterior distributions of the lens mass and distance are presented in 
Figures~\ref{fig:six} and \ref{fig:seven}, respectively.  Table~\ref{table:seven} 
summarizes the estimated host ($M_{\rm h}$) and planet ($M_{\rm p}$) masses, the 
distance to the planetary system, and the projected separation ($a_\perp$) between 
the planet and its host.  The table also lists the probabilities that the lens is 
located in the Galactic disk ($p_{\rm disk}$) or in the bulge ($p_{\rm bulge}$).  
For events with degenerate solutions, we present the projected separations 
corresponding to each solution. In the table, the median values are given as 
representative estimates, with the 16th and 84th percentiles of the Bayesian 
posteriors adopted as the lower and upper uncertainty limits.

Bayesian analyses indicate that all planets in the lens systems are giant planets, 
with masses comparable to or exceeding that of Jupiter in the Solar System. In each 
case, the host star is a main-sequence star with a mass lower than that of the Sun. 
For KMT-2024-BLG-0176, the lens system comprises a Jovian planet orbiting an M-dwarf 
host near the bottom of the main sequence. Such configurations are relatively rare, 
in contrast to the more common cases of Jupiter or super-Jupiter planets orbiting 
hosts of $\sim 0.6$–$0.7~M_\odot$.  The inferred locations of the planetary systems 
vary among the events: KMT-2024-BLG-0349L is most likely situated in the Galactic 
disk, whereas KMT-2024-BLG-0176L and KMT-2024-BLG-2087L are favored to lie in the 
bulge. For KMT-2024-BLG-1870L, the probabilities of a disk or bulge location are 
approximately equal.

\section{Summary and conclusion} \label{sec:six}

In summary, we have analyzed four newly discovered planetary microlensing events 
from the 2024 KMTNet season: KMT-2024-BLG-0176, KMT-2024-BLG-0349, KMT-2024-BLG-1870, 
and KMT-2024-BLG-2087. This work contributes to the ongoing systematic effort to build 
a complete and unbiased dataset of microlensing planets, which is essential for robust 
demographic studies.

Beyond the primary goal of comprehensive documentation, these four planets provide 
several important scientific insights. First, they highlight the morphological 
diversity of microlensing signals, ranging from a positive bump and an asymmetric 
peak to a dip and a caustic crossing. Second, all four are giant planets orbiting 
sub-solar-mass hosts, reinforcing the view that massive planets can form around 
low-mass stars. Third, their inferred locations span both the Galactic disk and 
bulge, thereby constraining models that connect planetary demographics with 
Galactic structure.
Although each individual detection may appear incremental, together they enrich 
the growing microlensing sample and offer methodological as well as astrophysical 
insights. In particular, they strengthen our understanding of planet formation 
beyond the snow line and further demonstrate the power of microlensing surveys 
to reveal planetary systems that are otherwise inaccessible.

\begin{acknowledgments}
Work by C.H. was supported supported by Chungbuk National University NUDP program (2025).
This research was supported by the Korea Astronomy and Space Science Institute under the R\&D 
program (Project No. 2025-1-830-05) supervised by the Ministry of Science and ICT.
This research has made use of the KMTNet system operated by the Korea Astronomy and Space Science 
Institute (KASI) at three host sites of CTIO in Chile, SAAO in South Africa, and SSO in Australia. 
Data transfer from the host site to KASI was supported by the Korea Research Environment Open NETwork 
(KREONET). 
J.C.Y., I.G.S., and S.J.C. acknowledge support from NSF Grant No. AST-2108414. 
H.Y. and W.Z. acknowledge support by the National Natural Science Foundation of
China (Grant No. 12133005). H.Y. acknowledge support by the China Postdoctoral Science Foundation (No.
2024M762938).
The MOA project is supported by JSPS KAKENHI Grant
Number JP24253004, JP26247023, JP23340064, JP15H00781, JP16H06287,
JP17H02871 and JP22H00153.
C.R. was supported by the Research fellowship of the Alexander von Humboldt Foundation.
The OGLE project has received funding from the Polish National Science
Centre grant OPUS-28 2024/55/B/ST9/00447 to AU. 
\end{acknowledgments}



\bibliographystyle{aasjournal}
\bibliography{pasp_refs}

\end{document}